\documentclass[proof]{pasj}

\begin{document} 

\title{Multi-band Photometry of Trans-Neptunian Objects in the Subaru Hyper Suprime-Cam Survey}

\author{Tsuyoshi \textsc{Terai}\altaffilmark{1}}
\altaffiltext{1}{Subaru Telescope, National Astronomical Observatory of Japan, National Institutes 
of Natural Sciences (NINS), 650 North A`ohoku Place, Hilo, HI 96720, USA}

\author{Fumi \textsc{Yoshida}\altaffilmark{2,3}}
\altaffiltext{2}{Planetary Exploration Research Center, Chiba Institute of Technology, 2-17-1 
Tsudanuma, Narashino, Chiba 275-0016, Japan}
\altaffiltext{3}{Department of Planetology, Graduate School of Science, Kobe University, 
Kobe 657-8501, Japan}

\author{Keiji \textsc{Ohtsuki}\altaffilmark{3}}

\author{Patryk \textsc{Sofia Lykawka}\altaffilmark{4}}
\altaffiltext{4}{School of Interdisciplinary Social and Human Sciences, Kindai University, 
Shinkamikosaka 228-3, Higashiosaka, Osaka 577-0813, Japan}

\author{Naruhisa \textsc{Takato}\altaffilmark{1}}

\author{Arika \textsc{Higuchi}\altaffilmark{5}}
\altaffiltext{5}{Department of Earth and Planetary Sciences, Faculty of Science, Tokyo Institute of 
Technology, Meguro, Tokyo 152-8551, Japan}

\author{Takashi \textsc{Ito}\altaffilmark{6}}
\altaffiltext{6}{National Astronomical Observatory of Japan, National Institutes of Natural Sciences 
(NINS), 2-21-1 Osawa, Mitaka, Tokyo 181-8588, Japan}

\author{Yutaka \textsc{Komiyama}\altaffilmark{6,7}}
\altaffiltext{7}{The Graduate University for Advanced Studies (SOKENDAI), 2-21-1 Osawa, Mitaka, 
Tokyo 181-8588, Japan}

\author{Satoshi \textsc{Miyazaki}\altaffilmark{6,7}}

\author{Shiang-Yu \textsc{Wang}\altaffilmark{8}}
\altaffiltext{8}{Institute of Astronomy and Astrophysics, Academia Sinica, P.O. Box 23-141, 
Taipei 10617, Taiwan}


\maketitle

\begin{abstract}
We present a visible multi-band photometry of trans-Neptunian objects (TNOs) observed by the Subaru 
Telescope in the framework of Hyper Suprime-Cam Subaru Strategic Program (HSC-SSP) from March in 
2014 to September in 2016.
We measured the five broad-band ($g$, $r$, $i$, $z$, and $Y$) colors over the wavelength range 
from 0.4~$\mu$m to 1.0~$\mu$m for 30 known TNOs using the HSC-SSP survey data covering 
$\sim$500~deg$^2$ of sky within $\pm30\degree$ of ecliptic latitude.
This dataset allows us to characterize the dynamical classes based on visible reflectance spectra 
as well as to examine the relationship between colors and the other parameters such as orbital 
elements.
Our results show that the hot classical and scattered populations share similar color 
distributions, while the cold classical population has a reflective decrease toward shorter 
wavelength below the $i$ band.
Based on the obtained color properties, we found that the TNO sample examined in the present work 
can be separated into two groups by inclination ($I$), the low-$I$ population consisting of cold 
classical objects and high-$I$ population consisting of hot classical and scattered objects.
The whole sample exhibits an anti-correlation between colors and inclination, but no significant 
correlation between colors and semi-major axis, perihelion distance, eccentricity, or absolute 
magnitude.
The color-inclination correlation does not seem to be continuous over the entire inclination range.
Rather, it is seen only in the high-$I$ population.
We found that the low- and high-$I$ populations are distinguishable in the $g-i$ vs. eccentricity 
plot, but four high-$I$ objects show $g-i$ colors similar to those of the low-$I$ population.
If we exclude these four objects, the high-$I$ objects show a positive correlation between $g-i$ 
and eccentricity and a negative correlation between $g-i$ and inclination with high significance 
levels.
\end{abstract}

\section{Introduction}\label{sec1}

Observational and theoretical studies of trans-Neptunian objects (TNOs) in the last quarter century 
have revolutionized not only our understanding of the outermost part of the solar system but also 
that of the formation of the solar system itself. 
The orbital distribution of TNOs provided a clear evidence for Neptune's outward migration 
\citep{Ml93}. 
In recent models for the origin of the solar system, all the four giant planets are thought to have 
experienced significant radial migration (e.g., \cite{Ts05,Wl11,NM12}), which caused injection of 
TNOs into the orbits of the giant planets; some of them were captured as irregular satellites of 
these planets or as Jupiter Trojan asteroids \citep{Mr05,Ns07,Ns13,Ns14}.
The scattered TNOs are thought to have reached even the outer asteroid belt region 
\citep{Lv09,Wl11}.
Orbital evolution of small bodies in the trans-Neptunian region has also been investigated based on 
such a model of giant planet migration \citep{Gm03,Lv08}.
Thus, observations of physical and dynamical properties of TNOs are expected to provide us with 
important and unique clues to clarify the processes of radial mixing of small bodies during the 
evolution of the solar system. 

Detailed information on the surface compositions and surface properties of TNOs can be acquired from 
spectroscopic observations (e.g., \cite{Ba08,Br12}).
However, due to their faintness, it is difficult to obtain their spectra with sufficient quality for 
a large number of TNOs. 
On the other hand, photometric observations using broad-band filters allow us to constrain surface 
properties of a larger number of objects and to obtain datasets relevant for statistical analysis 
(e.g., \cite{Dl04,Ba05,HD02,Hn12,Dr07,Px08,Px15,FB12,Fr15}).
For example, previous photometric observations revealed that colors of TNOs in the visible 
wavelength show a wide distribution from neutral to very red values (e.g., \cite{Dr08}). 
On the other hand, analysis of near-infrared color-color diagrams shows clustering of TNOs around 
the solar colors indicating flat reflectance spectra, although some TNOs have bluish colors due to 
absorption of surface ices such as H$_2$O and CH$_4$.
However, near-infrared color data for TNOs are rather limited. 
As for the correlation between colors and orbital elements, it is known that a population of the 
classical TNOs (e.g., \cite{LM07,Gl08}) with low inclination are covered by reddish surface 
(e.g., \cite{Dr08}).

Several mechanisms have been proposed to explain the observed colors of TNOs. 
Models based on a primordial origin assume that the surface properties reflect the radial distance 
from the Sun where the object formed \citep{Lw72}, while evolutionary models propose that subsequent 
evolution of TNOs such as impacts and/or space weathering could explain TNOs' color diversity 
\citep{Dr08,WB16,Sk17}. 
Thus, investigation of color distributions of TNOs and their correlations with other parameters such 
as orbital elements is expected to provide us useful and unique clues to reveal physical 
conditions of the planetesimal disk in the early stage of the solar system and/or history of its 
dynamical evolution.

In the present work, we examine color distribution of TNOs based on imaging data obtained through 
the Hyper Suprime-Cam Subaru Strategic Program (HSC-SSP) 
survey\footnote{http://hsc.mtk.nao.ac.jp/ssp/} \citep{Ah17}. 
The Hyper Suprime-Cam (HSC) is an optical imaging camera installed at the prime focus of the Subaru 
Telescope, and offers the widest field of view on existing 8--10~m class telescopes 
(\cite{My12}; Miyazaki et al. 2017, in preparation).
The HSC-SSP survey allows us to obtain unbiased high-quality photometric data even for faint objects 
such as TNOs. 
As part of our ongoing project based on this survey, in the preset work, we will examine color 
distributions of known TNOs, which were obtained with five broad-band filters ($g$, $r$, $i$, $z$, 
and $Y$). 
We will describe data used in the preset work in Section~2. 
Results of our data analysis are presented in Section~3, and are further discussed in Section~4. 
Our conclusions are summarized in Section~5.

\section{Data}\label{sec2}

\subsection{HSC-SSP survey}\label{sec2-1}

We use the imaging data obtained in the HSC-SSP survey.
The HSC is a prime focus camera for the 8.2-m Subaru Telescope, which consists of 116 2k~$\times$~4k 
Hamamatsu fully depleted CCDs (104 for science, 8 for focus monitoring, and 4 for auto guiding) and 
has a 1.5$\degree$ diameter field-of-view (FOV) with a pixel scale of 0$\farcs$168 
(\cite{My12}; Miyazaki et al. 2017, in preparation; Komiyama et al. 2017, in preparation).
The HSC-SSP project is a multi-band imaging survey with $g$, $r$, $i$, $z$, $Y$ broad-band and four 
narrow-band filters, covering 1400~deg$^2$ of the sky for 300 nights over 5--6 years from March 
2014 (Aihara et al. 2017, in preparation).
As of the first public data release on February 28, 2017, the HSC-SSP data taken between March 2014 
and November 2015 over a total of 61.5~nights has been 
released\footnote{https://hsc-release.mtk.nao.ac.jp} \citep{Ah17}. 
The exposure times of a single shot are 150--300~sec in the $g$ and $r$ bands and 200--300~sec in
the $i$, $z$, and $Y$ bands.
Note that the original $r$- and $i$-band filters of HSC ( ``HSC-r" and ``HSC-i" ) were replaced with 
new ones ( ``HSC-r2" and ``HSC-i2" ) in Jul 2016 and Feb 2016, respectively, but we regard the 
previous and present filters as the same performance because the transmission curves for the old and 
new ones are similar.

\begin{figure}
 \begin{center}
  \includegraphics[width=80mm]{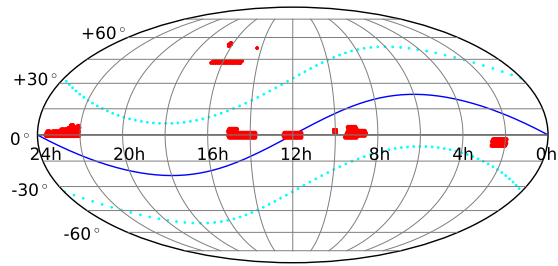} 
 \end{center}
\caption{
The sky map in equatorial coordinate covered by the Hyper Suprime-Cam Subaru Strategic Program 
survey between March 2014 and September 2016.
The solid and dotted curves show the ecliptic latitudes of $0\degree$ and $\pm30\degree$, 
respectively.
}\label{fig01}
\end{figure}

In this paper, we perform photometric investigation of known TNOs using the HSC-SSP data
obtained between March 2014 and September 2016.
The area covered by this dataset is $\sim$600~deg$^2$ with the five broad-band filters, as 
shown in Figure~\ref{fig01}, which includes $\sim$500~deg$^2$ of fields suitable for our detection 
of TNOs located within 30$\degree$ from the ecliptic plane.
The typical seeing size was 0.5--0.9~arcsec in all the broad-bands.

\subsection{Data reduction}\label{sec2-2}

The data is processed with {\tt hscPipe}, which is the HSC data reduction/analysis pipeline 
developed by the HSC collaboration team (Bosch et al. 2017, in preparation) based on the Large 
Synoptic Survey Telescope (LSST) pipeline software \citep{Iv08,Ax10,Jr15}.
First, a raw image is reduced by CCD-by-CCD procedures including bias subtraction, linearity 
correction, flat-fielding, artifacts masking, and background subtraction.
Next, the pipeline detects sources and determines the World Coordinate System (WCS) and  
zero-point magnitude of the corrected data by matching to the Pan-STARRS~1 (PS1) 3$\pi$ catalog 
\citep{Tn12,Sc12,Mg13}.
Then, the centroids, shapes, and fluxes of detected sources are measured with several different
algorithms.
We use the sinc aperture flux \citep{BL13} with 12~pixel (i.e., $\sim$2.0~arcsec) radius aperture 
for photometry of TNOs, which is the same configuration for estimating the zero-point magnitude.
Note that the zero-points are translated from PS1 into the native HSC system by a color term
(Kawanomoto et al. 2017, in preparation).
Lastly, the pipeline generates source catalogs describing the measurement values and flags of
detected objects in each CCD.
The source catalogs and corrected images are stored in the HSC database, and can be queried by 
using the Catalog Archive Server (CAS) 
search\footnote{https://hsc-release.mtk.nao.ac.jp/datasearch/} and quarried by using the Data 
Archive System (DAS)\footnote{DAS Quarry (https://hsc-release.mtk.nao.ac.jp/das\_quarry/) and DAS 
Search (https://hsc-release.mtk.nao.ac.jp/das\_search/pdr1/) are available for accessing the HSC-SSP 
image data.}, respectively (Takata et al. 2017, in preparation).

\subsection{Object sample}\label{sec2-3}

As of the end of 2016, more than 1700 TNOs have been discovered.
The daily ephemeris of each object was retrieved from the Minor Planet \& Comet Ephemeris Service 
website\footnote{http://www.minorplanetcenter.net/iau/MPEph/MPEph.html} managed by the Minor 
Planet Center.
We searched for known TNOs with coordinates located within the area of the HSC-SSP data at the 
acquisition date, and checked if there was a detected source corresponding to each of those objects 
in the source catalogs using the sub-hourly ephemeris.
The identified source was checked by visual inspection.
If the source is judged not to be a TNO (e.g., a star, galaxy, artifact) or has some kind of 
problem such as located at the edge of the image or too close to a very bright object, it is 
excluded from the color measurement.
In addition, we selected objects measured in at least four broad-bands as the targets of this
study.
Based on the above criteria we finally found 30 TNOs suitable for the present multi-band color 
investigation in the HSC-SSP data.
Table~\ref{tab01} shows these objects and their acquisition dates in each band.
Twenty eight objects were measured in all the five broad-bands, and the other two objects 
(2014~DL$_{143}$ and 2002~PD$_{155}$) were measured in the four bands.

\begin{longtable}{lcl}
\caption{
Measured TNOs and their acquisition dates in each filter.
The number of shots in each date is also displayed in parentheses. 
}\label{tab01}
\hline
\ \ Object \ \ \ \ \ \ \ \ \ \ \ & \ \ \ \ Filter \ \ \ \  & \ UT Date \\ 
\endfirsthead
\hline
\ \ Object \ \ \ \ \ \ \ \ \ \ \ & \ \ \ \ Filter \ \ \ \  & \ UT Date \\ 
\hline 
\endhead
\hline
\endfoot
\hline
\endlastfoot
\hline
1994 ES$_{2}$             & $g$ &  2015-03-25 (1), 2016-04-04 (6)  \\ 
                          & $r$ &  2016-04-08 (1), 2016-06-05 (2)  \\ 
                          & $i$ &  2015-03-22 (1), 2016-04-02 (8)  \\ 
                          & $z$ &  2015-03-16 (1), 2016-06-04 (8)  \\ 
                          & $Y$ &  2015-03-29 (1)  \\ 
(60454) 2000 CH$_{105}$   & $g$ &  2015-03-25 (5)  \\ 
                          & $r$ &  2015-03-18 (4)  \\ 
                          & $i$ &  2015-03-22 (1), 2015-03-20 (5)  \\ 
                          & $z$ &  2015-03-16 (4), 2015-05-23 (3)  \\ 
                          & $Y$ &  2015-03-29 (5)  \\ 
2001 DC$_{106}$           & $g$ &  2015-03-25 (2)  \\ 
                          & $r$ &  2015-03-18 (3)  \\ 
                          & $i$ &  2015-03-20 (2)  \\ 
                          & $z$ &  2015-03-16 (3)  \\ 
                          & $Y$ &  2015-03-29 (3)  \\ 
2001 DD$_{106}$           & $g$ &  2015-03-25 (4), 2016-04-04 (1)  \\ 
                          & $r$ &  2015-03-18 (3), 2016-04-08 (1)  \\ 
                          & $i$ &  2015-03-22 (1), 2015-03-20 (4)  \\ 
                          & $z$ &  2015-03-16 (6), 2015-05-23 (1)  \\ 
                          & $Y$ &  2015-03-29 (4), 2016-02-12 (1), 2016-04-15 (1)  \\ 
2002 CT$_{154}$           & $g$ &  2015-03-25 (1), 2016-04-04 (4)  \\ 
                          & $r$ &  2015-03-18 (1), 2016-04-08 (2)  \\ 
                          & $i$ &  2015-03-20 (1), 2016-04-02 (6)  \\ 
                          & $z$ &  2015-03-16 (1), 2016-06-04 (5)  \\ 
                          & $Y$ &  2015-03-29 (1), 2016-04-15 (4)  \\ 
2002 CY$_{154}$           & $g$ &  2015-03-25 (2)  \\ 
                          & $r$ &  2015-03-18 (4)  \\ 
                          & $i$ &  2015-03-20 (6)  \\ 
                          & $z$ &  2015-05-23 (4)  \\ 
                          & $Y$ &  2015-03-29 (3), 2015-05-19 (2)  \\ 
2002 FV$_{6}$             & $g$ &  2015-03-25 (3)  \\ 
                          & $r$ &  2015-03-18 (2)  \\ 
                          & $i$ &  2015-03-20 (2), 2015-05-21 (2)  \\ 
                          & $z$ &  2015-05-23 (2)  \\ 
                          & $Y$ &  2015-03-29 (2)  \\ 
2002 PD$_{155}$           & $g$ &  2014-10-01 (3), 2016-09-06 (1)  \\ 
                          & $r$ &  2014-09-22 (2), 2016-09-07 (1)  \\ 
                          & $i$ &  2014-09-22 (3)  \\ 
                          & $z$ &  2014-10-01 (5)  \\ 
(434194) 2003 FK$_{127}$  & $g$ &  2016-04-04 (3)  \\ 
                          & $r$ &  2016-06-05 (1)  \\ 
                          & $i$ &  2016-04-02 (6)  \\ 
                          & $z$ &  2016-06-04 (2)  \\ 
                          & $Y$ &  2016-04-15 (4)  \\ 
2003 FL$_{127}$           & $g$ &  2016-04-04 (2)  \\ 
                          & $r$ &  2016-04-08 (4), 2016-06-05 (3)  \\ 
                          & $i$ &  2016-03-04 (2)  \\ 
                          & $z$ &  2016-03-12 (5), 2016-06-01 (4)  \\ 
                          & $Y$ &  2016-02-03 (3)  \\ 
(120178) 2003 OP$_{32}$   & $g$ &  2016-07-07 (2)  \\ 
                          & $r$ &  2016-06-11 (3), 2016-09-02 (3)  \\ 
                          & $i$ &  2016-07-05 (4), 2016-08-30 (1), 2016-08-28 (1)  \\ 
                          & $z$ &  2016-06-04 (3), 2016-06-01 (1), 2016-07-12 (3)  \\ 
                          & $Y$ &  2016-08-09 (8), 2016-07-31 (4)  \\ 
(183963) 2004 DJ$_{64}$   & $g$ &  2016-04-04 (6)  \\ 
                          & $r$ &  2016-06-05 (1)  \\ 
                          & $i$ &  2016-04-02 (8)  \\ 
                          & $z$ &  2016-06-04 (6)  \\ 
                          & $Y$ &  2016-04-15 (7)  \\ 
2004 DM$_{71}$            & $g$ &  2016-04-04 (5)  \\ 
                          & $r$ &  2016-06-05 (3)  \\ 
                          & $i$ &  2016-04-02 (7)  \\ 
                          & $z$ &  2016-06-04 (3)  \\ 
                          & $Y$ &  2016-04-15 (2)  \\ 
2005 GX$_{186}$           & $g$ &  2016-04-04 (7)  \\ 
                          & $r$ &  2016-06-05 (5)  \\ 
                          & $i$ &  2016-04-02 (8)  \\ 
                          & $z$ &  2016-06-04 (5)  \\ 
                          & $Y$ &  2016-04-15 (3)  \\ 
(145452) 2005 RN$_{43}$   & $g$ &  2014-09-18 (2), 2015-07-14 (5)  \\ 
                          & $r$ &  2014-09-22 (2), 2015-07-15 (5)  \\ 
                          & $i$ &  2014-11-23 (2), 2015-07-21 (5)  \\ 
                          & $z$ &  2014-09-28 (1), 2015-07-22 (3), 2015-08-20 (1)  \\ 
                          & $Y$ &  2014-09-18 (2), 2015-07-23 (5)  \\ 
2008 SO$_{266}$           & $g$ &  2015-10-14 (3)  \\ 
                          & $r$ &  2015-10-10 (4)  \\ 
                          & $i$ &  2015-11-10 (5)  \\ 
                          & $z$ &  2015-11-06 (1), 2016-01-12 (2)  \\ 
                          & $Y$ &  2015-10-20 (7), 2015-10-06 (1)  \\ 
2012 VR$_{113}$           & $g$ &  2014-10-01 (1), 2014-11-18 (3), 2015-10-14 (2)  \\ 
                          & $r$ &  2014-11-18 (2), 2015-10-10 (2)  \\ 
                          & $i$ &  2014-09-22 (1), 2014-11-28 (2), 2014-11-23 (1)  \\ 
                          & $z$ &  2014-10-01 (2), 2015-11-06 (3)  \\ 
                          & $Y$ &  2014-09-18 (1), 2014-11-28 (5), 2015-10-20 (3)  \\ 
2013 QO$_{95}$            & $g$ &  2014-10-01 (3), 2014-11-18 (4), 2015-10-14 (4)  \\ 
                          & $r$ &  2014-11-18 (4), 2015-10-10 (4)  \\ 
                          & $i$ &  2014-09-22 (6), 2015-01-21 (1)  \\ 
                          & $z$ &  2014-10-01 (6), 2014-11-25 (1), 2014-11-21 (2)  \\ 
                          & $Y$ &  2014-09-18 (4), 2014-11-21 (2), 2016-01-09 (5)  \\ 
2013 QP$_{95}$            & $g$ &  2014-10-01 (6), 2014-11-18 (6), 2015-10-14 (1)  \\ 
                          & $r$ &  2014-11-28 (4), 2014-11-18 (9), 2015-10-10 (2)  \\ 
                          & $i$ &  2014-09-22 (8), 2014-11-23 (11)  \\ 
                          & $z$ &  2014-10-01 (6), 2014-11-25 (4), 2014-11-21 (7), 2015-11-06 (4)  \\ 
                          & $Y$ &  2014-09-18 (7), 2014-11-21 (14), 2016-01-09 (2)  \\ 
2013 RD$_{98}$            & $g$ &  2014-10-01 (4), 2014-11-25 (1), 2014-11-18 (4), 2015-10-14 (3)  \\ 
                          & $r$ &  2014-11-18 (5), 2015-10-10 (2)  \\ 
                          & $i$ &  2014-09-22 (5), 2014-11-23 (6)  \\ 
                          & $z$ &  2014-10-01 (2), 2014-11-25 (2), 2014-11-21 (5), 2016-01-12 (3)  \\ 
                          & $Y$ &  2014-09-18 (1), 2015-10-06 (2)  \\ 
2014 DL$_{143}$           & $g$ &  2016-04-04 (6)  \\ 
                          & $r$ &  2016-06-05 (5)  \\ 
                          & $i$ &  2016-04-02 (2)  \\ 
                          & $Y$ &  2016-04-15 (2), 2016-06-28 (2)  \\ 
2014 GS$_{53}$            & $g$ &  2015-05-17 (7), 2016-04-04 (6)  \\ 
                          & $r$ &  2015-07-13 (2), 2016-03-09 (3)  \\ 
                          & $i$ &  2015-05-21 (3), 2016-04-09 (8)  \\ 
                          & $z$ &  2015-07-12 (7), 2016-03-12 (5)  \\ 
                          & $Y$ &  2014-03-25 (2), 2015-05-26 (5), 2016-03-15 (7), 2016-03-14 (2)  \\ 
2014 GX$_{53}$            & $g$ &  2015-03-25 (1), 2015-05-17 (2)  \\ 
                          & $r$ &  2015-03-18 (1), 2015-05-15 (2)  \\ 
                          & $i$ &  2014-03-28 (7), 2015-03-22 (4), 2015-03-20 (2)  \\ 
                          & $z$ &  2015-03-29 (1), 2015-05-23 (5), 2015-05-13 (1)  \\ 
                          & $Y$ &  2014-03-25 (8), 2015-03-29 (1), 2015-05-19 (7)  \\ 
2014 NB$_{66}$            & $g$ &  2015-10-14 (1), 2016-07-07 (5)  \\ 
                          & $r$ &  2016-07-03 (3)  \\ 
                          & $i$ &  2016-07-05 (2), 2016-08-30 (2)  \\ 
                          & $z$ &  2016-07-29 (3)  \\ 
                          & $Y$ &  2016-07-31 (6)  \\ 
(483002) 2014 QS$_{441}$  & $g$ &  2014-11-25 (3)  \\ 
                          & $r$ &  2014-11-28 (2), 2014-11-18 (1)  \\ 
                          & $i$ &  2014-11-23 (2), 2015-01-21 (5), 2015-08-11 (4)  \\ 
                          & $z$ &  2014-11-25 (5), 2015-01-16 (4), 2015-08-20 (11), 2016-08-01 (4)  \\ 
                          & $Y$ &  2014-11-28 (3), 2015-08-09 (11)  \\ 
2014 TU$_{85}$            & $g$ &  2014-11-18 (3)  \\ 
                          & $r$ &  2014-11-18 (3)  \\ 
                          & $i$ &  2014-09-22 (5), 2014-11-23 (5), 2015-01-21 (2)  \\ 
                          & $z$ &  2014-10-01 (1), 2014-11-25 (2), 2014-11-21 (1), 2015-01-16 (1)  \\ 
                          & $Y$ &  2014-09-18 (1)  \\ 
2014 UF$_{224}$           & $g$ &  2014-10-01 (4), 2014-11-18 (10), 2015-10-14 (5)  \\ 
                          & $r$ &  2014-11-28 (5), 2014-11-18 (10), 2015-10-10 (5)  \\ 
                          & $i$ &  2014-09-22 (5), 2014-11-23 (11), 2015-11-14 (1), 2016-02-09 (2)  \\ 
                          & $z$ &  2014-10-01 (3), 2014-11-25 (8), 2014-11-21 (7), 2015-11-06 (5), 2016-01-12 (9)  \\ 
                          & $Y$ &  2014-11-21 (6)  \\ 
2014 WZ$_{509}$           & $g$ &  2015-03-25 (1)  \\ 
                          & $r$ &  2015-03-18 (1)  \\ 
                          & $i$ &  2015-01-21 (8)  \\ 
                          & $z$ &  2015-01-16 (6), 2015-11-06 (4)  \\ 
                          & $Y$ &  2015-01-27 (6), 2015-10-20 (1), 2016-01-09 (1)  \\ 
2015 FM$_{345}$           & $g$ &  2015-05-17 (5), 2016-04-04 (3)  \\ 
                          & $r$ &  2015-07-13 (1), 2016-04-08 (1)  \\ 
                          & $i$ &  2015-05-21 (1), 2016-04-09 (8)  \\ 
                          & $z$ &  2015-07-12 (8), 2016-03-12 (3)  \\ 
                          & $Y$ &  2015-05-26 (4), 2016-03-15 (4)  \\ 
2015 QT$_{11}$            & $g$ &  2015-10-14 (3)  \\ 
                          & $r$ &  2014-11-18 (1), 2015-10-10 (4)  \\ 
                          & $i$ &  2014-11-23 (2), 2016-02-09 (1)  \\ 
                          & $z$ &  2014-11-25 (1)  \\ 
                          & $Y$ &  2015-10-06 (2)  \\ 
\end{longtable}

The semi-major axis  ($a$), eccentricity ($e$), and inclination ($I$) of our target objects are 
listed in Table~\ref{tab02}.
According to the Deep Ecliptic Survey (DES) online 
classification\footnote{http://www.boulder.swri.edu/~buie/kbo/desclass.html} (Buie et al.), 
the sample includes four TNOs located at mean motion resonances (MMR) with Neptune, three in the 
3:2~MMR (2003~FL$_{127}$, 2008~SO$_{266}$, and 2013~RD$_{98}$) and one in the 2:1~MMR 
(2012~VR$_{113}$).
Next, we divided the non-resonant TNOs in our sample into two dynamical classes: classical and 
scattered TNOs.
Based on the classification scheme presented by \authorcite{LM07} (\yearcite{LM07}; see also
\cite{Gl08}), the former objects have orbits with semi-major axis of 37~au~$<$~$a$~$<$~47.5~au and 
perihelion distance $q$~$>$~37~au, while the latter ones have orbits with 30~au~$<$~$q$~$<$~37~au.
Using the above definition, we classified 19 objects as classical TNOs and 7 objects as scattered 
TNOs.
There was no object corresponding to the detached population in our sample.

\begin{figure}
 \begin{center}
  \includegraphics[width=160mm]{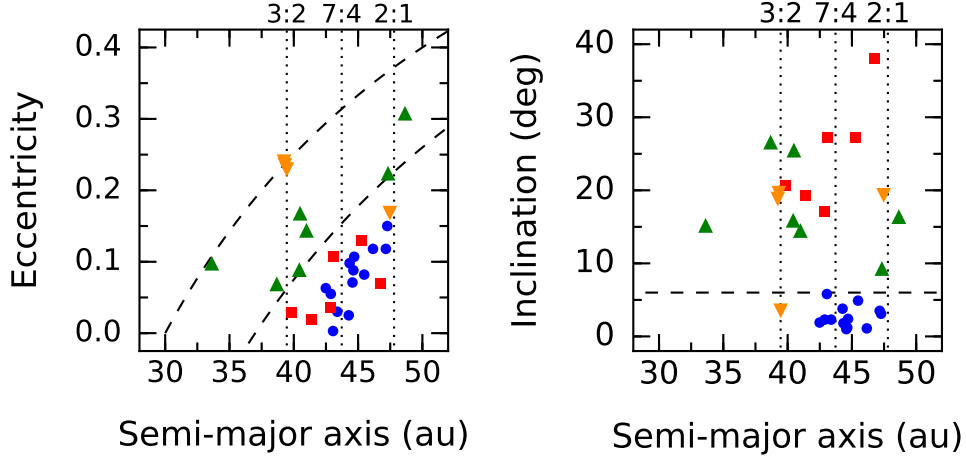} 
 \end{center}
\caption{
Semi-major axis vs. eccentricity (left) and semi-major axis vs. inclination (right) plots of our 
TNO sample.
Our 30 target TNOs are shown according to distinct dynamical classes: cold classical (circles), 
hot classical (squares), scattered (triangles), and resonant objects (inverse triangles), 
respectively.
The vertical dotted lines show the locations the of 3:2, 7:4, and 2:1 mean motion resonances with 
Neptune.
The dashed curves in the left panel show perihelion distances of 30~au and 37~au.
The dashed curve in the right panel shows an inclination of $6\degree$.
}\label{fig02}
\end{figure}

Furthermore, the classical TNOs contain a swarm of objects with inclination smaller than 6$\degree$, 
which is generally called "cold classical TNOs".
We divided the classical TNOs into two sub-populations with a boundary of inclination 
$I$~=~6$\degree$, 13 cold ($I$~$<$~6$\degree$) and 6 hot ($I$~$>$~6$\degree$) objects.
The $a$~vs.~$e$ and $a$~vs.~$I$ plots of our samples with the above classification are shown in 
Figure~\ref{fig02}. 

\begin{longtable}{lccrcccccc}
\caption{Semi-major axis ($a$), eccentricity ($e$), inclination ($I$), dynamical classification, and 
measured absolute magnitudes ($H$) in the $g$, $r$, $i$, $z$, and $Y$ bands.
}\label{tab02}
\hline
\ \ \ Object & $a$  & $e$ & $I$ \ \ \ & Class & $H_g$ & $H_r$ & $H_i$ & $H_z$ & $H_Y$ \\
             & (au) &     & (deg)     &       & (mag) & (mag) & (mag) & (mag) & (mag) \\
\endfirsthead
\hline
\ \ \ Object & $a$  & $e$ & $I$ \ \ \ & Class & $H_g$ & $H_r$ & $H_i$ & $H_z$ & $H_Y$ \\
             & (au) &     & (deg)     &       & (mag) & (mag) & (mag) & (mag) & (mag) \\
\hline 
\endhead
\hline
\endfoot
\hline
\endlastfoot
\hline
1994~ES$_{2}$   & 46.158 & 0.12 &  1.1 & Cold classical & 8.60~$\pm$~0.39 & 7.54~$\pm$~0.25 & 7.17~$\pm$~0.24 & 6.67~$\pm$~0.51 & 8.29~$\pm$~1.74 \\
2000~CH$_{105}$ & 44.636 & 0.09 &  1.2 & Cold classical & 7.45~$\pm$~0.21 & 6.40~$\pm$~0.14 & 6.13~$\pm$~0.11 & 6.04~$\pm$~0.11 & 5.97~$\pm$~0.27 \\
2001~DC$_{106}$ & 42.488 & 0.06 &  1.9 & Cold classical & 7.45~$\pm$~0.11 & 6.39~$\pm$~0.12 & 5.86~$\pm$~0.08 & 5.64~$\pm$~0.11 & 5.81~$\pm$~0.23 \\
2001~DD$_{106}$ & 44.338 & 0.10 &  1.8 & Cold classical & 8.53~$\pm$~0.34 & 7.56~$\pm$~0.10 & 7.06~$\pm$~0.22 & 6.90~$\pm$~0.28 & 6.70~$\pm$~0.58 \\
2002~CT$_{154}$ & 47.159 & 0.12 &  3.5 & Cold classical & 7.72~$\pm$~0.20 & 7.00~$\pm$~0.17 & 6.30~$\pm$~0.19 & 6.40~$\pm$~0.15 & 6.13~$\pm$~0.38 \\
2002~CY$_{154}$ & 44.564 & 0.07 &  1.0 & Cold classical & 7.86~$\pm$~0.27 & 6.74~$\pm$~0.12 & 6.29~$\pm$~0.12 & 6.25~$\pm$~0.34 & 5.51~$\pm$~0.45 \\
2002~FV$_{6}$   & 47.267 & 0.15 &  3.1 & Cold classical & 7.94~$\pm$~0.31 & 6.30~$\pm$~0.12 & 6.21~$\pm$~0.17 & 6.09~$\pm$~0.36 & 5.55~$\pm$~0.39 \\
2002~PD$_{155}$ & 43.051 & 0.00 &  5.8 & Cold classical & 8.22~$\pm$~0.28 & 7.31~$\pm$~0.74 & 6.96~$\pm$~0.13 & 7.18~$\pm$~0.26 &       n/a       \\
2003~FK$_{127}$ & 42.878 & 0.06 &  2.3 & Cold classical & 7.95~$\pm$~0.16 & 6.61~$\pm$~0.11 & 6.57~$\pm$~0.12 & 6.04~$\pm$~0.47 & 6.48~$\pm$~0.40 \\
2003~FL$_{127}$ & 39.461 & 0.23 &  3.5 & Resonant       & 7.72~$\pm$~0.17 & 6.50~$\pm$~0.12 & 5.93~$\pm$~0.22 & 5.91~$\pm$~0.21 & 5.66~$\pm$~0.16 \\
2003~OP$_{32}$  & 43.086 & 0.11 & 27.2 & Hot classical  & 4.11~$\pm$~0.09 & 3.79~$\pm$~0.14 & 3.63~$\pm$~0.11 & 3.67~$\pm$~0.09 & 3.68~$\pm$~0.08 \\
2004~DJ$_{64}$  & 44.698 & 0.11 &  2.4 & Cold classical & 7.61~$\pm$~0.17 & 7.15~$\pm$~0.12 & 6.63~$\pm$~0.15 & 6.88~$\pm$~0.59 & 6.20~$\pm$~0.27 \\
2004~DM$_{71}$  & 43.383 & 0.03 &  2.3 & Cold classical & 8.26~$\pm$~0.19 & 7.29~$\pm$~0.14 & 7.02~$\pm$~0.11 & 7.02~$\pm$~0.39 & 7.83~$\pm$~1.14 \\
2005~GX$_{186}$ & 44.273 & 0.03 &  3.8 & Cold classical & 7.61~$\pm$~0.21 & 6.46~$\pm$~0.12 & 6.58~$\pm$~0.15 & 6.17~$\pm$~0.24 & 6.11~$\pm$~0.60 \\
2005~RN$_{43}$  & 41.379 & 0.02 & 19.3 & Hot classical  & 4.26~$\pm$~0.06 & 3.48~$\pm$~0.04 & 3.18~$\pm$~0.04 & 3.02~$\pm$~0.07 & 2.99~$\pm$~0.06 \\
2008~SO$_{266}$ & 39.236 & 0.24 & 18.8 & Resonant       & 7.18~$\pm$~0.07 & 6.40~$\pm$~0.09 & 5.85~$\pm$~0.10 & 5.78~$\pm$~0.12 & 5.60~$\pm$~0.13 \\
2012~VR$_{113}$ & 47.466 & 0.17 & 19.3 & Resonant       & 7.04~$\pm$~0.12 & 6.37~$\pm$~0.05 & 6.15~$\pm$~0.10 & 6.09~$\pm$~0.20 & 5.96~$\pm$~0.30 \\
2013~QO$_{95}$  & 39.822 & 0.03 & 20.6 & Hot classical  & 7.79~$\pm$~0.31 & 6.58~$\pm$~0.10 & 6.40~$\pm$~0.35 & 6.11~$\pm$~0.15 & 5.84~$\pm$~0.47 \\
2013~QP$_{95}$  & 40.488 & 0.17 & 25.5 & Scattered      & 7.62~$\pm$~0.28 & 7.17~$\pm$~0.10 & 6.90~$\pm$~0.09 & 6.70~$\pm$~0.15 & 6.70~$\pm$~0.54 \\
2013~RD$_{98}$  & 39.315 & 0.24 & 19.6 & Resonant       & 7.75~$\pm$~0.17 & 7.20~$\pm$~0.13 & 7.02~$\pm$~0.27 & 6.81~$\pm$~0.36 & 6.45~$\pm$~0.33 \\
2014~DL$_{143}$ & 47.332 & 0.22 &  9.3 & Scattered      & 6.65~$\pm$~0.08 & 6.14~$\pm$~0.09 & 5.65~$\pm$~0.10 &       n/a       & 5.66~$\pm$~0.31 \\
2014~GS$_{53}$  & 33.622 & 0.10 & 15.2 & Scattered      & 7.34~$\pm$~0.09 & 6.31~$\pm$~0.16 & 5.80~$\pm$~0.09 & 5.57~$\pm$~0.12 & 5.48~$\pm$~0.13 \\
2014~GX$_{53}$  & 40.988 & 0.14 & 14.5 & Scattered      & 5.97~$\pm$~0.11 & 5.14~$\pm$~0.03 & 5.02~$\pm$~0.10 & 4.81~$\pm$~0.11 & 4.68~$\pm$~0.10 \\
2014~NB$_{66}$  & 45.476 & 0.08 &  4.9 & Cold classical & 5.89~$\pm$~0.19 & 5.66~$\pm$~0.12 & 5.13~$\pm$~0.12 & 4.69~$\pm$~0.10 & 5.05~$\pm$~0.28 \\
2014~QS$_{441}$ & 46.774 & 0.07 & 38.0 & Hot classical  & 5.58~$\pm$~0.14 & 5.31~$\pm$~0.09 & 5.28~$\pm$~0.06 & 5.30~$\pm$~0.09 & 5.46~$\pm$~0.33 \\
2014~TU$_{85}$  & 48.647 & 0.31 & 16.4 & Scattered      & 8.85~$\pm$~0.38 & 8.10~$\pm$~0.26 & 8.07~$\pm$~0.25 & 7.81~$\pm$~0.38 & 7.53~$\pm$~0.42 \\
2014~UF$_{224}$ & 45.273 & 0.13 & 27.2 & Hot classical  & 7.45~$\pm$~0.25 & 7.11~$\pm$~0.17 & 6.86~$\pm$~0.26 & 6.64~$\pm$~0.34 & 6.44~$\pm$~0.29 \\
2014~WZ$_{509}$ & 40.431 & 0.09 & 15.9 & Scattered      & 5.80~$\pm$~0.09 & 5.34~$\pm$~0.09 & 5.10~$\pm$~0.05 & 4.96~$\pm$~0.20 & 5.04~$\pm$~0.15 \\
2015~FM$_{345}$ & 42.862 & 0.04 & 17.1 & Hot classical  & 7.16~$\pm$~0.30 & 6.44~$\pm$~0.09 & 5.72~$\pm$~0.14 & 5.76~$\pm$~0.23 & 5.67~$\pm$~0.41 \\
2015~QT$_{11}$  & 38.676 & 0.07 & 26.6 & Scattered      & 8.97~$\pm$~0.17 & 8.32~$\pm$~0.16 & 8.56~$\pm$~0.64 & 8.00~$\pm$~0.40 & 7.44~$\pm$~0.26 \\

\end{longtable}

\subsection{Color measurements}\label{sec2-4}

The apparent magnitude ($m$) of the individual sources in each band is estimated from the 
photometric zero-point determined by the {\tt hscPipe} processing, and is converted into the 
absolute magnitude $H$ of TNOs as
\begin{equation}
H = m - 5 \log ( R \cdot \Delta ) - \beta \cdot \alpha,
\label{eq01}
\end{equation}
where $R$ and $\Delta$ are the heliocentric and geocentric distances in au, $\alpha$ is the
solar phase angle, and $\beta$ is the phase coefficient.

\citet{Rb07} examined the plot of $B$- and $I$-band phase coefficients obtained from 18 TNOs
and 7 Centaurs (see Figure~3 of their paper).
Although there are several outliers in TNOs, most of the objects have comparable values between the
two bands.
\citet{Sc09} also presented the differences in the phase curve slopes between $B$ and $I$ bands
for 52 icy bodies, and showed that in most cases they are as small as 
$\lesssim$~$\pm$0.02~mag~deg$^{-1}$.
These data imply that the phase coefficient of TNOs can almost be regarded as wavelength 
independent.
We assume a constant value of $\beta$~=~0.11~mag~deg$^{-1}$ \citep{AC16} for all the bands.

We obtain the object absolute magnitude $H_x$ in a band $x$ (either of $g$, $r$, $i$, $z$, $Y$) 
by averaging the individual absolute magnitudes over all the epochs, $H_{x,j}$ ($j$~=~1, 2,..., 
$N_x$, where $N_x$ is the number of the $x$-band data).
The uncertainty of $H_x$, $\sigma_{H_x}$, is calculated as 
\begin{equation}
 \sigma_{H_x} = \sqrt{ \frac{ \sum^{N_x}_{j=1} \sigma^2_{H_{x,j}} }{N_x^2}
                     + \frac{ \sum^{N_x}_{j=1} ( H_{x,j} - H_x )^2 }{N_x - 1} },
\label{eq02}
\end{equation}
where $\sigma_{H_{x,j}}$ is the error of individual magnitude given by the square root of the square 
sum of the photometric error derived from flux measurement uncertainty and the zero-point error.
The second term in the square root, the unbiased variance of $H_{x,j}$, is larger than the first 
term derived from the individual errors in most sources, suggesting that the $H_{x,j}$ dispersions 
are dominated by brightness variation with rotation of the TNO.

We estimate the color from the difference between $H_x$ values, e.g., $H_g - H_r$ for the $g-r$ 
color.
The deviation of the $H_x$ value from the mean magnitude level of the lightcurve ($\Delta H_x$) 
could cause an additional uncertainty in the color measurement.
According to the analysis by \citet{Df09}, the mean rotation period and lightcurve amplitude of 
TNOs are 6.95~hr and 0.25~mag, respectively.
Using a Monte Carlo method, we generated synthetic lightcurves assuming a sinusoidal brightness 
fluctuation with 6.95~hr period, 0.25~mag amplitude, and random initial phase angles, and computed 
the standard deviation ($\sigma_{\Delta H_x}$) of $\Delta H_x$ values at the actual acquisition 
epochs of each object/band.
Then, the total $H_x$ uncertainty is obtained by 
$\sqrt{ \sigma_{H_x}^2 + \sigma_{\Delta H_x}^2 }$.
The $H_x$ magnitudes (in the AB magnitude system) and their total errors of all the sample objects 
are shown in Table~\ref{tab02}.

\begin{figure}
 \begin{center}
  \includegraphics[width=160mm]{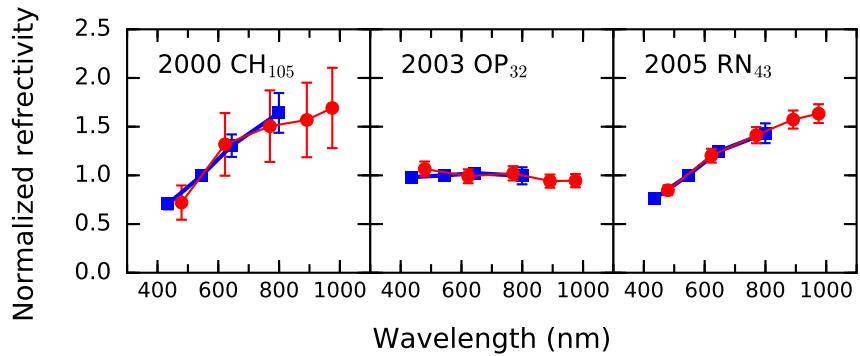} 
 \end{center}
\caption{
Normalized reflectance spectra of the objects whose colors were previously measured and compiled in 
the ``Minor Bodies in the Outer Solar System" (MBOSS) database \citep{HD02,Hn12}.
The circles and squares show the color data obtained by the present work from the HSC-SSP survey 
($g$, $r$, $i$, $z$, and $Y$ bands) and those compiled in MBOSS ($B$, $V$, $R$, and $I$ bands), 
respectively.
}\label{fig03}
\end{figure}

Three of our sample objects have been observed with multi-band imaging by previous studies: 
(60454) 2000~CH$_{105}$ \citep{Px04,Bn11}, (120178) 2003~OP$_{32}$ \citep{Pr10,Pr13,Rb08}, 
and (145452) 2005~RN$_{43}$ \citep{DM09,Pr13}, which are also listed in the ``Minor Bodies in the 
Outer Solar System" (MBOSS) database \citep{HD02,Hn12}.
The reflectance spectra of these objects obtained from the MBOSS color data and our measurements are 
shown in Figure~\ref{fig03}.
The solar colors were given by \citet{Hl06} and were corrected into the HSC band system with color 
term conversion (see Table~\ref{tab03}).

\begin{table}
\tbl{
Colors of the Sun.
}{
 \centering
  \begin{tabular}{ccr}
    \hline
    Color & SDSS\footnotemark[$*$] & HSC \\
    \hline
    $g-r$ & 0.45 &    0.40 \\
    $r-i$ & 0.12 &    0.13 \\
    $i-z$ & 0.04 &    0.04 \\
    $z-Y$ & ---  & $-$0.01 \\
    \hline
  \end{tabular}}\label{tab03}
  \begin{tabnote}
    \footnotemark[$*$] \citet{Hl06}
  \end{tabnote}
\end{table}

In the range of wavelengths where the two datasets overlap, we confirmed that the two spectra agree 
well with each other for all of the three objects, indicating that our measurements have sufficient 
accuracy.
Note that our color data with the wavelength range from 0.4~$\mu$m to 1.0~$\mu$m is useful to 
complement the spectral coverage between $I$ ($\sim$0.8~$\mu$m) and $J$ ($\sim$1.25~$\mu$m) bands.
To our knowledge, except the three aforementioned objects, the visible colors of our sample objects 
were measured for the first time by our present study.

\section{Results}\label{sec3}

\begin{figure}
 \begin{center}
  \includegraphics[width=160mm]{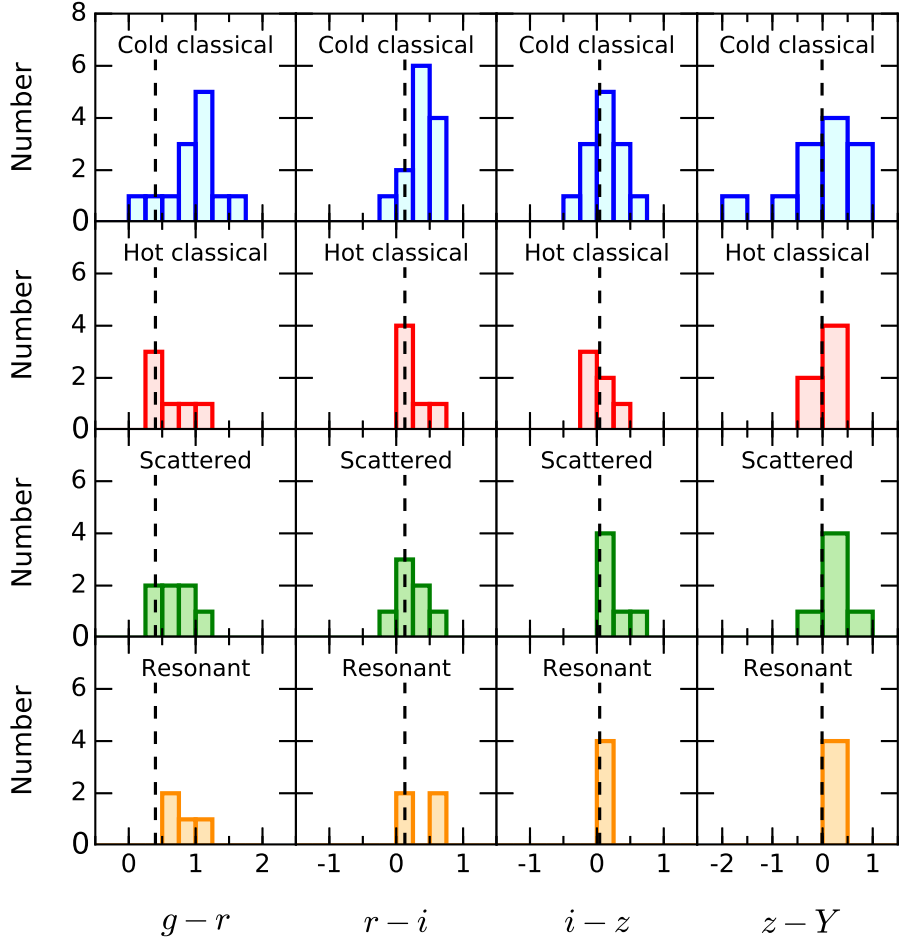} 
 \end{center}
\caption{
Color distributions of our target TNOs according to their dynamical classes.
From top to bottom, the rows show the cold classical, hot classical, scattered, and resonant 
populations.
The dashed lines show the solar color.
}\label{fig04}
\end{figure}

First, we examine colors derived from the absolute magnitudes in two adjacent bands.
Figure~\ref{fig04} shows the histogram plots of the sample objects in the $g-r$, $r-i$, $i-z$, and 
$z-Y$ colors, separately displayed for each dynamical population.
The dashed lines represent the solar color.
This figure reveals the following properties:
(1)~The hot classical and scattered populations have similar color distributions with the peak close 
to the solar color in all band pairs.
(2)~The cold classical population exhibits a redder distribution compared to the hot classical and 
scattered populations in the $g-r$ and $r-i$ colors, while their $i-z$ and $z-Y$ colors are 
distributed around the solar color.
(3)~The resonant objects concentrate close to the solar color in the $i-z$ 
and $z-Y$ bands, but their colors apparently range from neutral to red in $g-r$ and $r-i$ colors.

\begin{figure}
 \begin{center}
  \includegraphics[width=160mm]{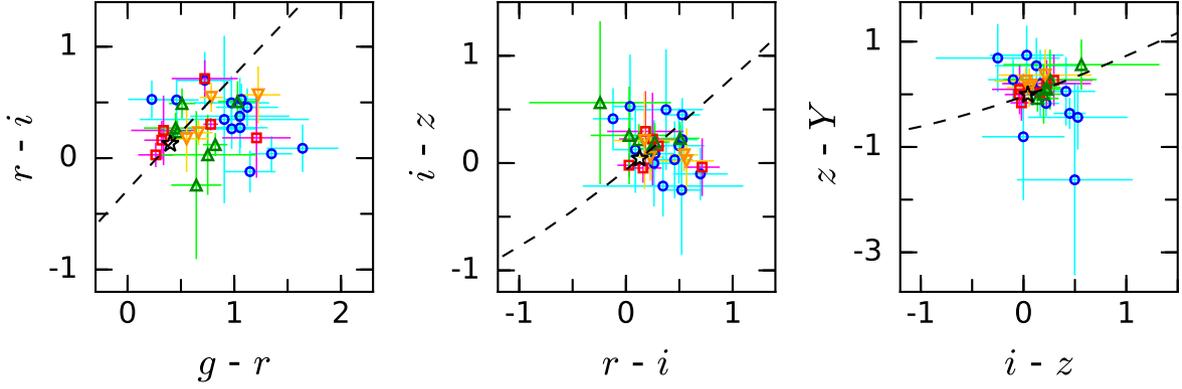} 
 \end{center}
\caption{
Color-color diagrams of our target TNOs according to distinct dynamical classes: 
cold classical (circles), hot classical (squares), scattered (triangles), and resonant objects 
(inverse triangles).
The star symbol shows the solar colors.
The dashed curves show the reddening lines (see text).
}\label{fig05}
\end{figure}

\begin{figure}
 \begin{center}
  \includegraphics[width=80mm]{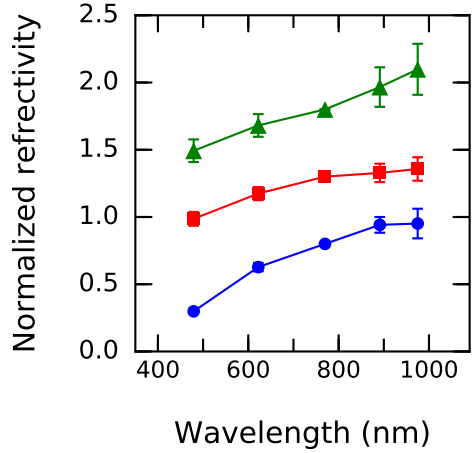} 
 \end{center}
\caption{
Average reflectance spectra of the cold classical (circles), hot classical (squares), and scattered 
(triangles) populations, which have been normalized at the $i$ band and vertically shifted for 
clarity.
}\label{fig06}
\end{figure}

Figure~\ref{fig05} illustrates the color-color diagrams of the sample objects, separately displayed 
for each dynamical population.
The star symbol shows the solar colors.
The dashed line indicates a color track given from linear reflectance spectra called 
``reddening line" \citep{HD02}.
In the $i-z$~vs.~$z-Y$ plot, most of the objects are located on the upper right side from the solar 
color along the reddening line, suggesting that the typical reflectance spectra of TNOs are flat or 
linearly increasing with wavelength between the $i$ and $Y$ bands regardless of the dynamical 
population.
Such a distribution is also seen for the hot classical and scattered populations in the 
$g-r$~vs.~$r-i$ and $r-i$~vs.~$i-z$ plots.
On the other hand, the cold classical objects show a significant deviation to the right side of the 
reddening line in the $g-r$~vs.~$r-i$ and potentially $r-i$~vs.~$i-z$ plots, indicating a steeper 
slope in the short wavelength range.
These properties can also be confirmed in the average reflectance spectra shown in 
Figure~\ref{fig06}.
The spectra of the hot classical and scattered populations are approximated by linear slopes
over all the five bands, but the reflectance spectra of the cold classical population show a steep 
drop at the short wavelength edge.

By comparing the color plots of the dynamical populations in Figures~\ref{fig04}, 
\ref{fig05}, and \ref{fig06}, it appears that the hot classical and scattered populations have 
similar color distributions, while the cold classical population exhibits a different spectral 
property especially in the short wavelength range.
In the following, we investigate the similarity of the color distributions among the populations 
using two statistical tests based on \citet{Pr02}.

First, the Student's $t$-test is used to evaluate whether the means of the measurement variable
(i.e., colors) are significantly different between two datasets.
The statistic $t$ for the unequal-variance $t$-test is computed by
\begin{equation}
t = \frac{ \overline{x_A} - \overline{x_B} }{ \sqrt{ {\rm Var}(x_A)/N_A + {\rm Var}(x_B)/N_B } },
\label{eq03}
\end{equation}
where $x_A$ and $x_B$ in the present case are the color distributions in the two datasets, and 
$\overline{x}$ and Var($x$) are their means and variances, respectively.
The probability for the null hypothesis that the two distributions are the same ($P_t$) is computed
from the incomplete beta function,
\begin{equation}
I_x(a,b) = \frac{\Gamma(a+b)}{\Gamma(a)\Gamma(b)} \int^x_0 t^{a-1} (1-t)^{b-1} dt,
\label{eq04}
\end{equation}
where $\Gamma$ is the gamma function, with $x = \nu/(\nu+t^2)$, $a = \nu/2$, and $b = 1/2$.
$\nu$ is the degree of freedom given by
\begin{equation}
\nu = \frac{ \left[ {\rm Var}(x_A)/N_A + {\rm Var}(x_B)/N_B \right]^2 }
           { \left[ {\rm Var}(x_A)/N_A \right]^2/(N_A-1)
           + \left[ {\rm Var}(x_B)/N_B \right]^2/(N_B-1) }.
\label{eq05}
\end{equation}

Second, the Kolmogorov-Smirnov (KS) test is used to verify if two populations originate from the 
same parent population. 
For this purpose, the KS statistic $D$ is given by
\begin{equation}
D = \max_{-\infty < x < \infty} | S_A(x) - S_B(x) |,
\label{eq06}
\end{equation}
where $S_A(x)$ and $S_B(x)$ are the cumulative probability distributions of a color $x$ of 
the two datasets.
The probability for the null hypothesis that the two distributions are the same is expressed by
\begin{equation}
P_{\rm KS} = Q_{\rm KS} \left( \left[ \sqrt{N_e} + 0.12 + 0.11/\sqrt{N_e} \right] D \right), 
\label{eq07}
\end{equation}
where $N_e = N_A N_B / (N_A + N_B)$ is the effective number of data points, and $Q_{\rm KS}$ is a 
function given by
\begin{equation}
Q_{\rm KS}(\lambda) = 2 \sum^{\infty}_{j=1} (-1)^{j-1} e^{-2j^2\lambda^2}.
\label{eq08}
\end{equation}

We compared the $g-r$ and $r-i$ color distributions among the cold classical, hot classical, and 
scattered populations using the above two tests.
The results are presented in Table~\ref{tab04}.
Both $P_t$ and $P_{\rm KS}$ probabilities between the cold classical and other populations show 
small values ($\lesssim$~0.05) in the $g-r$ color, supporting to reject the null hypothesis.
In contrast, between the hot classical and scattered populations, both $P_t$ and $P_{\rm KS}$ are 
larger than 0.2 for the two colors examined, indicating that the color distributions are not 
significantly different for the two populations.

\begin{table}
\tbl{
Results of Student's $t$-test and Kolmogorov-Smirnov (KS) test.
}{
  \begin{tabular}{lccc}
    \hline
    Populations & Color & $P_t$\footnotemark[$*$] & $P_{\rm KS}$\footnotemark[$\dag$] \\ 
    \hline
    Cold classical vs. Hot classical & $g-r$ & 0.04 & 0.06 \\
                                     & $r-i$ & 0.57 & 0.28 \\
    Cold classical vs. Scattered     & $g-r$ & 0.01 & 0.03 \\
                                     & $r-i$ & 0.12 & 0.34 \\
    Hot classical vs. Scattered      & $g-r$ & 0.34 & 0.28 \\
                                     & $r-i$ & 0.48 & 0.95 \\
    \hline
  \end{tabular}}\label{tab04}
\begin{tabnote}
  \footnotemark[$*$]    Probabilities for the null hypothesis in the $t$-test\\
  \footnotemark[$\dag$] Probabilities for the null hypothesis in the KS test
\end{tabnote}
\end{table}

In addition, the two-dimensional KS test was applied to compare the $g-r$ vs. $r-i$ color
distributions among the three dynamical populations.
The statistic $D$ for this test is obtained from the maximum difference of the integrated 
probabilities in each of four natural quadrants around a given point.
The probability for the null hypothesis is expressed by
\begin{equation}
P_{\rm 2KS} = Q_{\rm KS} \left( \frac{ \sqrt{N_e} D }{ 1 + \sqrt{1-r_p^2} (0.25 - 0.75/\sqrt{N_e}) } 
\right),
\label{eq09}
\end{equation}
where $r_p$ is the Pearson's coefficient given by
\begin{equation}
r_p = \frac{ \sum^{N}_{j=1} (x_j - \overline{x}) (y_j - \overline{y}) }
           { \sqrt{ \sum^{N}_{j=1} (x_j - \overline{x})^2 } 
             \sqrt{ \sum^{N}_{j=1} (y_j - \overline{y})^2 } }
\label{eq10}
\end{equation}
for pairs of quantities ($x_j$, $y_j$), $j$ = 1, 2,..., $N$.
We obtained 
$P_{\rm 2KS}$ = 0.012 between the cold classical and hot classical populations,
$P_{\rm 2KS}$ = 0.004 between the cold classical and scattered populations, and
$P_{\rm 2KS}$ = 0.141 between the hot classical and scattered populations.
These results indicate that there is a significant difference of colors between the cold classical 
and other populations, while the color distributions are similar between the hot classical and 
scattered populations, as the previous tests suggested.

On the basis of the above analysis of color distributions for each dynamical population, our target 
TNOs can be divided into two groups at $I$~$\sim$~6$\degree$: the low-inclination component 
containing the cold classical objects and the high-inclination component containing the hot 
classical and scattered objects, hereinafter called ``low-$I$" and ``high-$I$" populations, 
respectively.
The resonant objects were excluded from the analysis below because their eccentricities and 
inclinations are likely to have changed after the capture into the resonances.
Both low- and high-$I$ populations consist of 13 objects.

\begin{figure}
 \begin{center}
  \includegraphics[width=80mm]{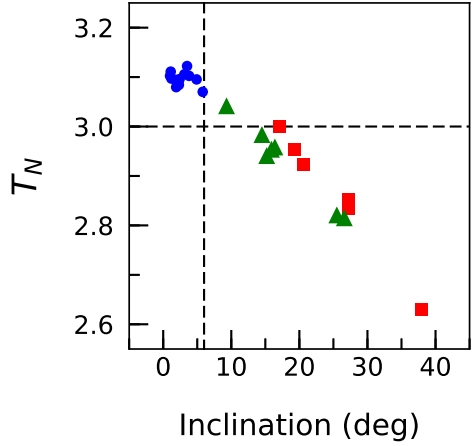} 
 \end{center}
\caption{
Inclination ($I$) vs. Tisserand parameter with respect to Neptune ($T_N$) plot of our target TNOs
according to their dynamical classes: cold classical (circles), hot classical (squares), and
scattered (triangles).
The vertical and horizontal dashed lines represent to $I$~=~6$\degree$ and $T_N$~=~3.0, 
respectively.
}\label{fig07}
\end{figure}

Following the above analysis of color distributions of the dynamical populations, we found that this 
grouping based on the inclinations matches reasonably well the division based on color properties.
In addition, as for the Tisserand parameter with respect to Neptune, 
$T_N$\footnote{$T_N = a_N/a + 2 \sqrt{ a ( 1 - e^2 )/a_N } \cos I^{\prime}$, where $a_N$ is 
Neptune's semi-major axis and $I^{\prime}$ is inclination with respect to the orbital plane of 
Neptune.}, the low-$I$ and high-$I$ populations have $T_N$~$>$~3.0 and $T_N$~$<$~3.0, 
respectively, except for one scattered object, 2014~DL$_{143}$ (see Figure~\ref{fig07}).
Thus, the division of our sample into the low- and high-$I$ populations is based on both physical 
(colors) and dynamical ($T_N$) grounds.

2014~DL$_{143}$ has a perihelion distance of 36.92~au close to the boundary between the classical 
and scattered TNOs, and inclination of 9.3$\degree$, which is by far smallest in the high-$I$ 
population.
This object could have intermediate dynamical characteristics between the cold/hot classical and 
scattered TNOs, while its $g-r$ color, 0.51~$\pm$~0.12, is closer to the color distribution peaks 
of the hot classical and scattered TNOs rather than that of the cold classical TNOs.
In this paper, 2014~DL$_{143}$ was classified into the high-$I$ population.

\begin{figure}
 \begin{center}
  \includegraphics[width=160mm]{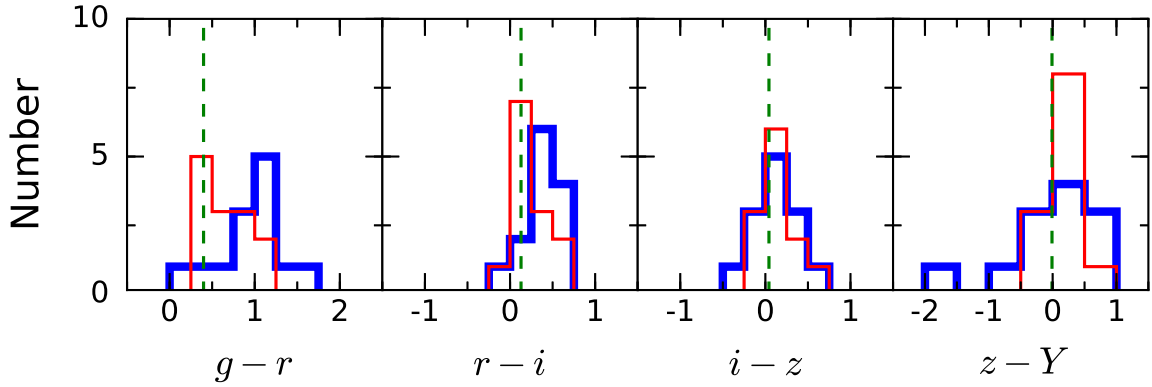} 
 \end{center}
\caption{
Color distributions of the low-$I$ (thick line) and high-$I$ (thin line) populations, defined by 
$I$~$<$~6$\degree$ and $I$~$>$~6$\degree$, respectively. 
The dashed lines show the solar color.
}\label{fig08}
\end{figure}

Figure~\ref{fig08} shows the color distributions of the low-$I$ and high-$I$ populations.
It is clear that the two populations differ in the $g-r$ and $r-i$ color distributions.
However, both populations are indistinguishable from each other in the $i-z$ and $z-Y$ color 
distributions.
The high-$I$ objects concentrate around the solar color in all the colors with slightly reddening 
with wavelength.
In contrast, most of the low-$I$ objects are remarkably redder than the solar color in the $g-r$ 
distribution, but the peak approaches the neutral in the long wavelength range.
This spectral distinction between these two populations may be caused by differences in their origin 
and/or evolution.
In the next section, we analyze the characteristics of the low-/high-$I$ populations as well as 
their correlations with other parameters.

\section{Discussion}\label{sec4}

\subsection{Correlations}\label{sec4-1}

\begin{figure}
 \begin{center}
  \includegraphics[width=160mm]{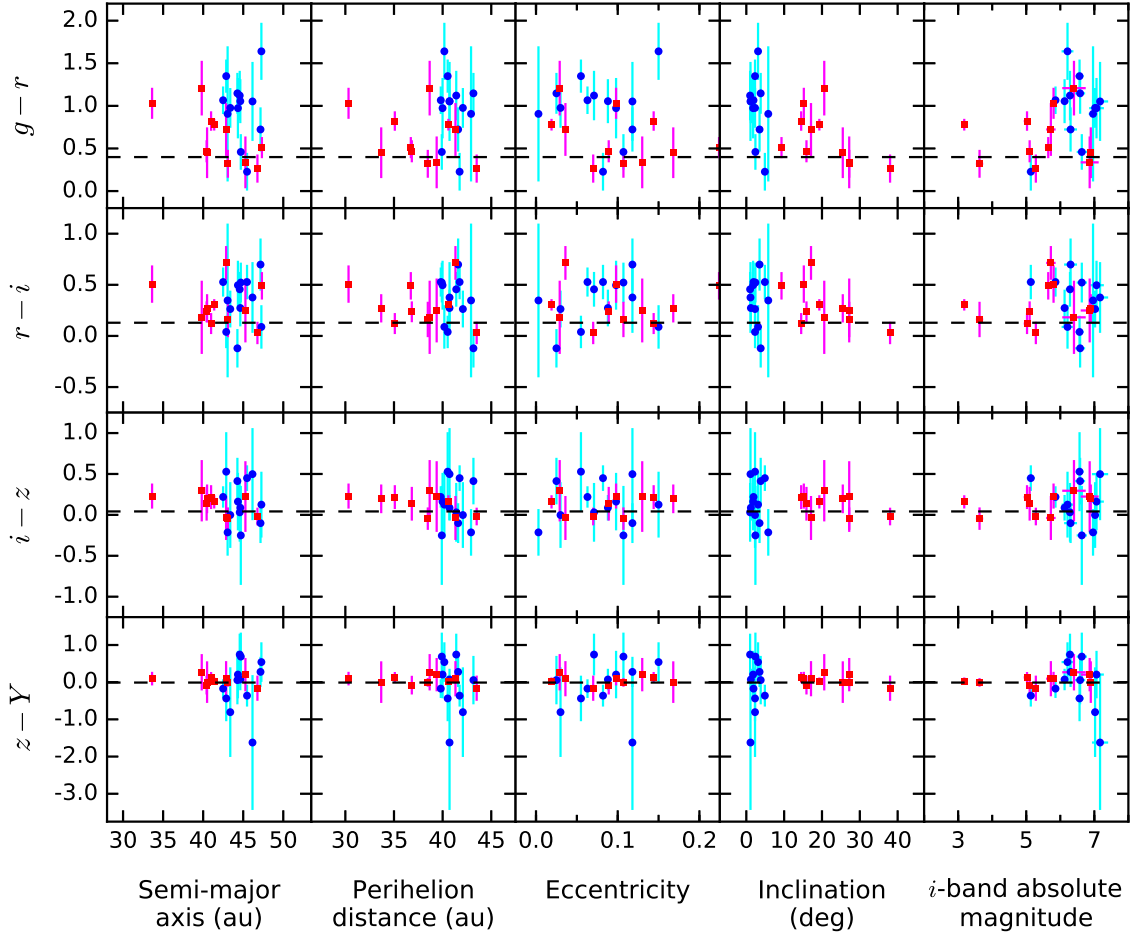} 
 \end{center}
\caption{
Colors vs. orbital elements and absolute magnitude of the low-$I$ (circles) and high-$I$ (squares) 
populations.
The dashed lines show the solar color.
}\label{fig09}
\end{figure}

Examination of the correlations between TNOs' colors and the other parameters such as orbital
elements could give us important clues on the origin and dynamical/chemical evolutions of TNOs.
In particular, comparison of such correlations may be useful to better understand the cause of 
similarity/difference in the color properties.
We assessed correlations of colors with the semi-major axis, perihelion distance, eccentricity, 
inclination, and absolute magnitude with the $i$ band.
The obtained colors vs. these parameters are displayed in Figure~\ref{fig09}.

Concerning the quality of the orbits of our sample, we referred to the JPL Small-Body Database 
Browser\footnote{http://ssd.jpl.nasa.gov/sbdb.cgi} for information regarding the precision of orbit 
determination of the individual objects.
As of March 2017, we found that 28 of them have sufficiently accurate orbits, but 2014~TU$_{85}$ 
and 2015~QT$_{11}$ have large orbital uncertainties in both semi-major axis and eccentricity 
($\Delta a$~$\sim$~22.2~au and $\Delta e$~$\sim$~0.58 for 2014~TU$_{85}$, $\Delta a$~$\sim$~0.85~au 
and $\Delta e$~$\sim$~0.36 for 2015~QT$_{11}$, where $\Delta a$ and $\Delta e$ are 1-$\sigma$ 
uncertainties of semi-major axis and eccentricity, respectively).
Therefore, these two objects were excluded from the analysis of correlations.

We applied the Spearman rank-order correlation coefficient $r_s$ to the selected sample objects to 
examine if the $g-r$ and $r-i$ colors are correlated with any of the above parameters.
For $N$ pairs of measurements ($x_j$, $y_j$), $r_s$ is given by
\begin{equation}
r_s = 1 - \frac{ 6 \sum^{N}_{j=1} (X_j - Y_j)^2 }{ N^3 - N },
\label{eq11}
\end{equation}
where $X_j$ and $Y_j$ are the ranks of $x_j$ and $y_j$, respectively.
$r_s$ is converted into Student's $t$ statistic by
\begin{equation}
t = r_s \sqrt{ \frac{ N - 2 }{ 1 - r_s^2 } },
\label{eq12}
\end{equation}
which is approximately distributed as the $t$ distribution with $\nu = N-2$.
The probability $P_t$ for the null hypothesis that the variables are uncorrelated is estimated from 
the incomplete beta function given by Equation~(\ref{eq04}).

Table~\ref{tab05} shows the results of the correlation tests.
We found significant correlations between colors and inclinations for all the samples.
Notably, the $g-r$ color exhibits strong anti-correlation with inclination ($P_t \sim 0.0003$), 
in agreement with previous studies (e.g., \cite{TB02,Dr08,Px08,Hn12}).
In contrast, the other parameters investigated are not significantly correlated with the colors.

\begin{table}
\tbl{
Significance of correlation between colors and other parameters.
}{
 \centering
  \begin{tabular}{lccrrrrr}
    \hline
    Sample & Color & $r_s$\footnotemark[$*$], $P_t$\footnotemark[$\dag$]
                   & $a$\footnotemark[$\ddag$] \  
                   & $q$\footnotemark[$\S$] \ 
                   & $e$\footnotemark[$\|$] \ 
                   & $I$\footnotemark[$\#$]
                   & $H_i$\footnotemark[$\**$] \\
    \hline
    All      & $g-r$ & $r_s$ & $-$0.15 &    0.10 & $-$0.22 &   $-$0.60 &    0.19 \\
             &       & $P_t$ &    0.73 &    0.49 &    0.08 & $\ll$0.01 &    0.10 \\
             & $r-i$ & $r_s$ & $-$0.07 &    0.02 &    0.08 &   $-$0.37 & $-$0.21 \\ 
             &       & $P_t$ &    0.33 &    0.10 &    0.37 &      0.02 &    0.09 \\
    \hline
    Low-$I$  & $g-r$ & $r_s$ & $-$0.31 & $-$0.27 &    0.07 &   $-$0.19 & $-$0.21 \\ 
             &       & $P_t$ &    0.14 &    0.84 &    0.22 &      0.57 &    0.63 \\
             & $r-i$ & $r_s$ &    0.03 & $-$0.41 &    0.46 &      0.05 & $-$0.35 \\ 
             &       & $P_t$ &    0.09 &    0.08 &    0.06 &      0.18 &    0.12 \\
    \hline
    High-$I$ & $g-r$ & $r_s$ & $-$0.39 & $-$0.31 & $-$0.08 &   $-$0.54 &    0.04 \\ 
             &       & $P_t$ &    0.07 &    0.12 &    0.27 &      0.02 &    0.15 \\
             & $r-i$ & $r_s$ & $-$0.39 & $-$0.31 & $-$0.08 &   $-$0.54 &    0.04 \\
             &       & $P_t$ &    0.53 &    0.13 &    0.03 &      0.06 &    0.72 \\
    \hline
  \end{tabular}}\label{tab05}
\begin{tabnote}
  \footnotemark[$*$]     Spearman rank-order correlation coefficient \\
  \footnotemark[$\dag$]  Probability for the null hypothesis of no correlation \\
  \footnotemark[$\ddag$] Semi-major axis \\
  \footnotemark[$\S$]    Perihelion distance \\
  \footnotemark[$\|$]    Eccentricity \\
  \footnotemark[$\#$]    Inclination \\
  \footnotemark[$\**$]   $i$-band absolute magnitude
\end{tabnote}
\end{table}

We also examined correlations for each of the low-/high-$I$ populations in the same manner.
The high-$I$ population exhibits a moderate correlation between the $g-r$ color and inclination, 
while the low-$I$ population does not show such a trend.
This may indicate that the correlation between color and inclination is not continuous across all 
data values, but is limited to the objects in the high-$I$ population.
If so, the low-/high-$I$ populations could represent distinct populations with different origins or 
dynamical evolutions.

\subsection{$g-i$ color}\label{sec4-2}

As discussed in Section~\ref{sec3}, Figure~\ref{fig08} illustrates that the low-/high-$I$ 
populations seem to have different $g-r$ and $r-i$ color distributions.
As suggested in previous works (e.g., \cite{WB16, WB17}), we take the $g-i$ color as a useful 
index to characterize the two populations.

\begin{figure}
 \begin{center}
  \includegraphics[width=80mm]{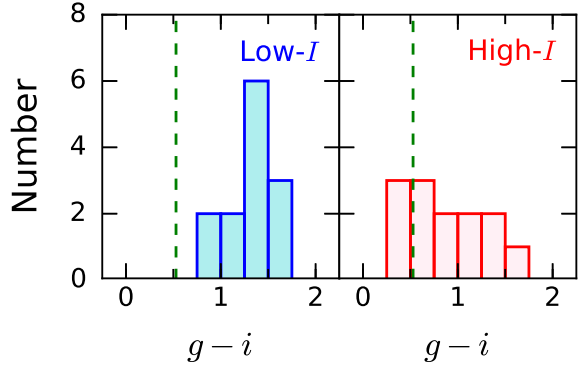} 
 \end{center}
\caption{
$g-i$ color distributions of the low-$I$ (left) and high-$I$ (right) populations.
The dashed lines show the solar color.
}\label{fig10}
\end{figure}

Figure~\ref{fig10} shows the $g-i$ color distributions for the two populations.
There is a clear distinction that the high-$I$ population exhibits a peak around the solar color
($0.25 \lesssim g-i \lesssim 0.75$) while the low-$I$ population has no corresponding objects
near the solar color.
The $g-i$ color of the high-$I$ distribution extends up to $g-i$~$\sim$~1.75, which corresponds to 
the higher end of the $g-i$ distribution for the low-$I$ population.
\citet{WB17} reported that the hot TNO population with $I \geq 5\degree$ has a bimodal distribution 
in $g-i$ color, ``red" (R; $g-i$~$\leq$~1.03) and ``very red" (VR; $g-i$~$\geq$~1.31) 
sub-populations.
The high-$I$ population of our data covers a similar range in $g-i$ color as in \citet{WB17}, but 
small number statistics prevent us from confirming bimodality.
We note that the peak location of their VR population, $g-i$~$\sim$~1.4, agrees with that of the 
low-$I$ distribution in our data.

\begin{figure}
 \begin{center}
  \includegraphics[width=80mm]{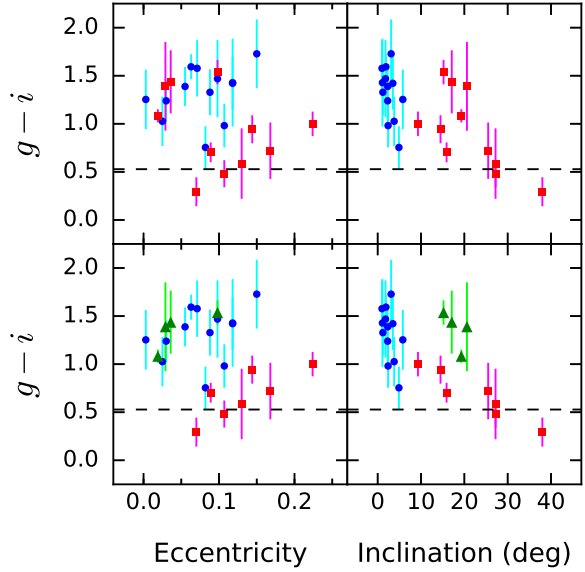} 
 \end{center}
\caption{
Top: $g-i$ color vs. eccentricity/inclination of the low-$I$ (circles) and high-$I$ (squares) 
populations.
Bottom: same as the top panels but the red high-$I$ objects are marked with triangles.
The dashed lines show the solar color.
}\label{fig11}
\end{figure}

Furthermore, we also found a notable relationship between the $g-i$ color and orbital elements.
Figure~\ref{fig11} illustrates the $g-i$ color vs. eccentricity and $g-i$ color vs. inclination for 
the low-/high-$I$ populations.
The $g-i$ color vs. eccentricity distribution seems to be divided into two swarms.
The first swarm consists of low eccentricity ($e \lesssim 0.1$) and red objects ($g-i \gtrsim 1.0$), 
and the second one consists of objects with relatively high eccentricity ($e \gtrsim 0.1$) and 
close to the neutral color.
The former is dominated by the low-$I$ population, while the latter is dominated by the high-$I$ 
population.

The $g-i$ vs. eccentricity plot also shows that four high-$I$ objects with low eccentricity 
(2005~RN$_{43}$, 2013~QO$_{95}$, 2014~GS$_{53}$, and 2015~FM$_{345}$) are located in the 
low-$e$/red swarm.
Intriguingly, these objects are concentrated at $a$~$<$~43~au and $I$~$<$~21$\degree$. 
Taken together with their redder $g-i$ colors, these results suggest that the high-$I$ population 
would consist of two distinct components, in agreement with the findings of \citet{WB17}. 
In particular, these four TNOs could be members of the VR TNOs, while the other high-$I$ TNOs 
would not belong to the VR group (furthermore, most of them are even less red than the mean $g-i$ 
color of the R TNOs).
If we focus the analysis solely on the latter group, after excluding these four objects, the $g-i$ 
distribution of the high-$I$ population exhibits a positive correlation with eccentricity and 
negative correlation with inclination with high significance levels ($P_t$~$\sim$~0.004 and 0.0005, 
respectively).
This may reflect the relationship among orbits, colors, and dynamical evolution of the 
neutral-color high-$I$ population, as well as possibly different origins/evolutions between the 
red and neutral high-$I$ objects.
These two sub-populations of high-$I$ objects probably form the color bimodality presented in 
\citet{WB17}, although the color distribution of the neutral high-$I$ objects disagree with that of 
the R TNOs.
While further data is required for detailed investigation, the presence of the red high-$I$ objects 
has potential for being an essential clue for understanding the cause of difference in reflectance 
spectra between low-$I$ and high-$I$ populations.

\subsection{Interpretation}\label{sec4-3}

Our analysis shows that (1) the hot classical and scattered populations have similar reflectance 
spectra with a constant (flat or reddish) slope in the wavelength range at least from 0.4~$\mu$m to 
1.0~$\mu$m; (2) the cold classical population exhibits distinctive spectra with a reflective 
decrease toward shorter wavelength below the $i$ band ($\sim$0.8~$\mu$m); (3) the high-$I$ 
population shows an anti-correlation between $g-r$/$r-i$ colors and inclination; (4) for the 
high-$I$ TNOs with less red color, there is a strong anti-correlation between $g-i$ color and 
inclination.
The depth, shape and wavelength range of this reflectance decay is potentially useful for 
elucidating the reddening mechanism as well as the factor of the color diversity of TNOs, which can 
provide a unique constraint on the origins and evolutions of these populations.

The difference in the color distributions among the dynamical classes and the color-inclination 
correlation have been noted in previous studies (e.g., \cite{Dr08,Hn12,Px15,WB17}).
Mainly two hypotheses have been proposed to explain these features, as we describe below.

\subsubsection{Collisional resurfacing}\label{sec4-3-1}

The diversity of colors would reflect different collisional evolutions experienced by the TNOs. 
While giant impacts may darken the surfaces of large TNOs \citep{Sk17}, it is often thought that 
impacts excavate surfaces reddened by space weathering \citep{Dr08}.
In particular, objects with dynamically excited orbits, i.e., with high eccentricity and/or 
inclination, can suffer more energetic collisions.
Such impacts can excavate the surface layers covered by irradiated reddish materials, thus exposing 
the subsurface fresh (neutral color) materials.
\citet{Hn12} reported an anti-correlation between color and the orbital 
excitation parameter $\varepsilon = \sqrt{e^2 + \sin^2 I}$ among the hot classical objects.
However, this parameter seems mostly dominated by the inclinations since most of the classical 
objects have moderately low eccentricities ($e \lesssim 0.2$).
The color-$\varepsilon$ correlation possibly reflects nothing but the color-inclination correlation.
In fact, we did not find a correlation between color and eccentricity in the low-$I$ population, 
which has a narrow range of inclinations.
This is inconsistent with the suggestion by \citet{TD03} that the colors of TNOs under collisional 
resurfacing should have a much stronger correlation with eccentricity than with inclination.
Therefore, our results do not seem to support their scenario.

\subsubsection{Cosmogonic origin}\label{sec4-3-2}

The diversity of colors would reflect the distinct formation environments. 
That is, the birthplaces of TNOs distributed across the protoplanetary disk in the early solar 
system.
Based on recent planetary migration models, the high-$I$ population would consist of objects 
transported from the inner regions of the disk due to migration of the giant planets, while the 
low-$I$ population would have formed near its present location (e.g., \cite{Lv08,Wl12,Fr14}).

The orbits of the latter objects would have allowed them to retain at least 
some of the volatile species such as CH$_3$OH, NH$_3$, CO$_2$, H$_2$S, and C$_2$H$_6$ ices on the 
surfaces (e.g., \cite{Br11,WB16}), which have been detected on comets and planetary satellites
\citep{Hd08,Cl13}.
These ices are considered to be able to create red material such as organics via irradiation by 
energetic particles and UV (e.g., \cite{Mr03,Br06}).
In contrast, the surfaces of objects formed in the inner regions with the temperature beyond the 
sublimation points of such ices maintain the original (neutral) color.
Assuming this hypothesis, the similarity of color distribution between the hot classical and
scattered population seems to indicate a common origin.
Furthermore, the close anti-correlation between $g-i$ color and inclination of the high-$I$
population with relatively high eccentricities and the existence of the reddish high-$I$ objects 
with low eccentricities could provide useful information about their dynamical evolution processes.

\section{Summary}

We measured the visible five-band colors of 30 known TNOs using the HSC-SSP survey data.
The object sample contains 13 cold classical, 6 hot classical, 7 scattered, and 4 resonant TNOs.

The color distributions of the hot classical and scattered populations have peaks close to the 
solar color in all the colors examined in this work, while the cold classical population shows 
deviations toward redder colors in the $g-r$ and $r-i$ distributions.
Statistical tests suggest that the hot classical and scattered populations share the same color 
property that the reflectance spectra are approximately linear.
On the other hand, the cold classical population deviates from the ``reddening line" in the 
$g-r$ vs. $r-i$ diagram, indicating that its reflectance spectra have a steep drop at the short 
wavelength edge.
This leads us to divide the samples (excluding resonant objects) into two groups, the low-$I$ 
population consisting of the cold classical objects and the high-$I$ population consisting of the 
hot classical and scattered objects.

The whole sample exhibits a significant anti-correlation between $g-r$/$r-i$ colors and inclination.
This correlation has also been detected in the high-$I$ population, but not in the low-$I$ 
population, suggesting that this relationship is probably not continuous over the entire inclination 
range.

The $g-i$ color is a useful index to characterize the low-/high-$I$ populations.
In agreement with the findings of \citet{WB17}, our high-$I$ population data support the bimodal 
distribution in $g-i$.
We also found that the sample can be separated into two swarms in the $g-i$ vs.~eccentricity plot, 
red/low-$e$ and neutral/high-$e$ groups.
Although most of the high-$I$ objects are contained in the neutral/high-$e$ group, four of them
are located in the red/low-$e$ region.
Excluding these objects, the $g-i$ color of the high-$I$ population exhibits a highly significant
positive correlation with eccentricity and negative correlation with inclination.
Our results showed that these color properties of TNOs can provide important and useful clues for 
a better understanding of the formation and evolution of small bodies in the outer solar 
system.

\begin{ack}
This study is based on data collected at Subaru Telescope and retrieved from the HSC data archive 
system, which are operated by Subaru Telescope and Astronomy Data Center, National Astronomical 
Observatory of Japan.

The Hyper Suprime-Cam (HSC) collaboration includes the astronomical communities of Japan and Taiwan, 
and Princeton University. 
The HSC instrumentation and software were developed by the National Astronomical Observatory of 
Japan (NAOJ), the Kavli Institute for the Physics and Mathematics of the Universe (Kavli IPMU), the 
University of Tokyo, the High Energy Accelerator Research Organization (KEK), the Academia Sinica 
Institute for Astronomy and Astrophysics in Taiwan (ASIAA), and Princeton University. 
Funding was contributed by the FIRST program from Japanese Cabinet Office, the Ministry of 
Education, Culture, Sports, Science and Technology (MEXT), the Japan Society for the Promotion of 
Science (JSPS), Japan Science and Technology Agency (JST), the Toray Science Foundation, NAOJ, Kavli 
IPMU, KEK, ASIAA, and Princeton University.

This paper makes use of software developed for the Large Synoptic Survey Telescope. 
We thank the LSST Project for making their code available as free software at http://dm.lsst.org.

The Pan-STARRS1 Surveys (PS1) have been made possible through contributions of the Institute for 
Astronomy, the University of Hawaii, the Pan-STARRS Project Office, the Max-Planck Society and its 
participating institutes, the Max Planck Institute for Astronomy, Heidelberg and the Max Planck 
Institute for Extraterrestrial Physics, Garching, The Johns Hopkins University, Durham University, 
the University of Edinburgh, Queen's University Belfast, the Harvard-Smithsonian Center for 
Astrophysics, the Las Cumbres Observatory Global Telescope Network Incorporated, the National 
Central University of Taiwan, the Space Telescope Science Institute, the National Aeronautics and 
Space Administration under Grant No. NNX08AR22G issued through the Planetary Science Division of 
the NASA Science Mission Directorate, the National Science Foundation under Grant No. AST-1238877, 
the University of Maryland, Eotvos Lorand University (ELTE), and the Los Alamos National Laboratory.

KO was supported by JSPS KAKENHI (No. 15H03716).
\end{ack}


\end{document}